# DMRG study of scaling exponents in spin-1/2 Heisenberg chains with dimerization and frustration

Manoranjan Kumar

Solid State and Structural Chemistry Unit, Indian Institute of Science, Bangalore 560012, India

S. Ramasesha

Solid State and Structural Chemistry Unit, Indian Institute of Science, Bangalore 560012, India; Jawaharlal Nehru Centre for Advanced Scientific Research, Jakkur, Bangalore 560064, India

Diptiman Sen

Centre for High Energy Physics, Indian Institute of Science, Bangalore 560012, India

Z.G. Soos

Department of Chemistry, Princeton University, Princeton, NJ 08544, USA

Abstract

In conformal field theory, key properties of spin-1/2 chains, such as the ground state energy per site and the excitation gap scale with dimerization $\delta$ as $\delta^\alpha$ with known exponents $\alpha$ and logarithmic corrections. The logarithmic corrections vanish in a spin chain with nearest ($J=1$) and next nearest neighbor interactions ($J_2$), for $J_{2c}=0.2411$. DMRG analysis of a frustrated spin chain with no logarithmic corrections yields the field theoretic values of $\alpha$, and scaling relation is valid up to the physically realized range, $\delta \sim 0.1$. However, chains with logarithmic corrections ($J_2<0.2411J$) are more accurately fit by simple power laws with different exponents for physically realized dimerizations. We show the exponents decreasing from approximately 3/4 to 2/3 for the spin gap and from approximately 3/2 to 4/3 for the energy per site and error bars in the exponent also decrease as $J_2$ approaches to $J_{2c}$.

PACs number: 75.10.Pq, 75.10.Jm

The linear Heisenberg antiferromagnet (HAF) of spin-1/2 sites is an important and unique model many-body system that is both experimentally realized (in organic and inorganic crystals[1]) and is theoretically amenable to exact solution. As discussed in Refs. 2 and 3, bozonization and conformal field theories have motivated recent theoretical interest in HAFs and dimerized spin chains and provide scaling laws for stabilization of the ground state energy per site as well as the magnitude of spin gap, as a function of the dimerization δ. However, the spin gap at experimentally realized dimerization, does not follow scaling[2,4]. Scaling theory leaves open the range of dimerization over which scaling results are reliable, while experiment requires HAFs with substantial dimerization that may be outside the range of scaling. Frustration in AF systems is another broad topic of current interest[5-10]. For example, a frustrated HAF has a second-neighbor exchange $J_2$ that yields a valence-bond solid at $J_2 = 1/2$. Since the scaling of the dimerization gap depends on $J_2$, an HAF with both dimerization and frustration[6] is well suited to study scaling exponents, logarithmic corrections and the dimerization range of scaling. The Hamiltonian with nearest-neighbor exchange $J = 1$, taken as the unit of energy, is

$$H(\delta, J_2) = \sum_n [1+\delta(-1)^n]\vec{s}_n \cdot \vec{s}_{n+1} + J_2 \vec{s}_n \cdot \vec{s}_{n+2} \qquad (1)$$

The parameters δ and $J_2 > 0$ describe dimerization and frustration, respectively. The regular HAF with $\delta = J_2 = 0$ in Eq. (1) has ground state (GS) energy per site[11] of $\varepsilon_0 = -\ln 2 + 1/4$ and vanishing singlet-triplet (ST) gap[12] $\Delta(0,0) = 0$. The ST gap, $\Delta(0,J_2)$, opens at $J_{2c} = 0.2411$, that marks the transition from magnetic to nonmagnetic ground state[13].

In this brief report, we use the density matrix renormalization group (DMRG) method[14] to find the ST gap and GS energy per site of the infinite chain. We analyze the results for $\delta \ll 1$ and $J_2 \leq J_{2c}$ as

$$\begin{aligned}\Delta(\delta, J_2) &= A\delta^\alpha \\ \varepsilon(\delta, J_2) - \varepsilon(0, J_2) &= -B\delta^\beta\end{aligned} \qquad (2)$$

The coefficients A, B and exponents α, β are functions of $J_2$. The exponents at $J_2 = 0$ are α = 2/3 and β = 4/3 according to bosonization and conformal field theory[3,15-19]. However, there are logarithmic corrections[3,15-19] so that the gap goes as $\delta^{2/3}/|\ln\delta|^{1/2}$ and the energy gain as $\delta^{4/3}/|\ln\delta|$. In a perturbation analysis of H(δ,0), Barnes, Riera and Tennnant (BRT)[4,20] found the ST gap to be accurately given by α = 3/4 and A = 2 over the entire range of δ from 0 to 1. This remarkably simple, accurate, but not exact expression showed that $\delta^{2/3}/|\ln\delta|^{1/2}$ is restricted to the surprisingly small range, δ < 0.02. Logarithmic corrections arise because H(δ,0) is not strictly scale invariant at δ = 0; a marginal operator is present that destroys scale invariance. The coefficient of the marginal operator decreases with increasing $J_2$ and vanishes at $J_{2c}$, where exponents α = 2/3 and β = 4/3 are then expected[6]. The dimerization range over which Eq. (2) is valid is given by $J_{2c}$. For $J_2$ = 0, Papenbrock et al also found a pure power law or a power law with field theoretic exponents and logarithmic corrections from their DMRG studies[4].

H(δ,$J_2$) conserves the total spin S, and the GS is a singlet, S = 0. Valence bond (VB) diagrams are a many-electron basis that conserves S and are best suited for exact diagonalization of small systems. DMRG calculations with fixed $M_S$ are highly accurate for large systems. The S=0 space is absent in the $M_S$ = 1 sector and hence allows computation of the ST gaps[6]. We computed ε(0,0) for open chains of N = 200 sites as half of the difference between the total energy for 198 and 200 sites, retaining 128, 200 and 300 density-matrix eigenvectors. All the three match to 8 digits. The sum of the eigenvalues of the discarded density matrix eigenvectors in each of these cases is approximately $10^{-15}$. We have further checked our numerical results using the finite system DMRG algorithm keeping 128 density matrix eigenvectors and doing finite system calculation at each chain length (before increasing the system size) up to 200 sites. We find that for $J_2$ = 0 and δ = 0 (the most difficult case due to the infinite correlation length for these parameter values[21]), the difference between the gaps in the finite and infinite system DMRG algorithms are in the fifth significant place at a chain length of 200 sites. All results below are based on infinite DMRG algorithm with 128 density matrix eigenvectors for open chains with even N from N = 100 to 200, and subsequent extrapolation to infinite N. The ground state energy per site at chain length, N is calculated ½[E(N)-E(N-2)], where E(N) is the energy of the lowest $M_S = 0$ at chain length, N. The spin gap at N is the energy difference between the lowest $M_S = 1$ state and the corresponding ground state. These quantities are plotted against 1/N and the thermodynamic values are obtained by extrapolation to infinite N using a fourth degree polynomial in (1/N). The extrapolated $\varepsilon(0,0)$ is < 2 x $10^{-8}$ J from the exact $\varepsilon_0$, while the extrapolated $\Delta(0,0)$ is less than $10^{-5}$ J. Comparable or improved convergence is expected for nonzero $\delta$ and $J_2$ with a finite ST gap, and is found on comparing to BRT[20] at $\delta = 1/3$, $J_2 = 0$. The range of $\delta$ extends ref. 20 down to $\delta = 0.001$. Direct targeting of spin state S improves the accuracy of the ST gap at small $\delta$ over previous works summarized in Table 3 of ref. 2.

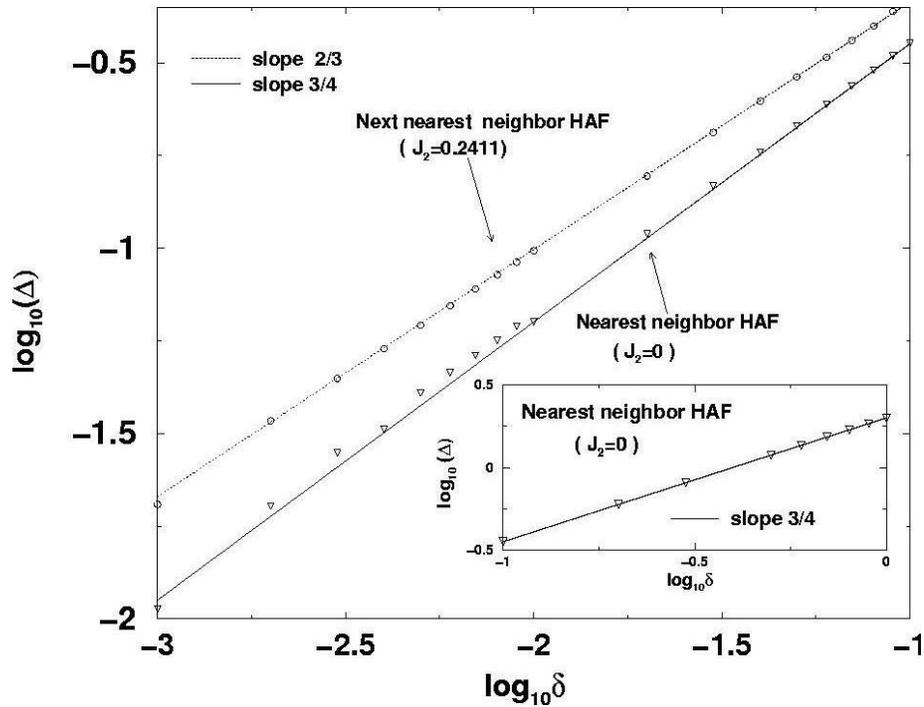

**Fig .1** Log-log plot of the ST gap $\Delta$ versus dimerization $\delta$ at $J_2 = J_{2c} = 0.2411$ and $J_2 = 0$. Scaling theory predicts a slope of 2/3 as found at $J_{2c}$ Inset extends the range of $\delta$ to 1 for $J_2 = 0$.

Table 1 lists representative values of extrapolated ST gaps and GS energies at $J_2 = 0$ and $J_{2c}$. Calculations were also performed at $J_2 = 0.05, 0.10, 0.15$ and $0.20$, as well as at other $\delta$. Figure 1 shows a log-log plot of the ST gaps between $\delta = 0.001$ and $0.10$ for $J_2 = 0$

(HAF) and $J_{2c}$. The inset extends the ST gap to $\delta = 1$. The HAF curve for the gap is the BRT relation,[4,20] $2\delta^{3/4}$, whose accuracy we confirm over three decades in $\delta$. A least squares fit between 0.001 and 0.10 returns an exponent of $\alpha = 0.750$. As discussed in Fig. 29 of ref. 2, small deviations from $\delta^{3/4}$ can be demonstrated. With A = 2 in Eq. (2) and $\Delta(\delta,J_2)$ in Table 1, the exponent $\alpha(J_2)$ is easily found at each $J_2$, and small deviations from 3/4 exceed the $10^{-5}$ accuracy of $\Delta(\delta,J_2)$. The BRT expansion[20] of the ST gap in a = (1-$\delta$)/(1+$\delta$) matches $\delta^{3/4}$ up to $a^2$ for a << 1, in the limit of weakly interacting dimers, but deviates in the next order. By contrast, the $J_{2c}$ gap has $\alpha = 2/3$ at least to $\delta = 0.10$, in agreement with scaling[3,15-19] and previous[6] DMRG results.

**Table 1.** Ground state energy per site, $\varepsilon(\delta,J_2)$, and singlet-triplet gap, $\Delta(\delta,J_2)$, of chain, Eq. (1), with dimerization $\delta$ and frustration $J_2$ for two special values = 0 (HAF) and 0.2411 ($J_{2c}$).

| Dimerization $\delta$ | $\varepsilon(\delta,0)$ | $\Delta(\delta,0)$ | $\varepsilon(\delta,J_{2c})$ | $\Delta(\delta,J_{2c})$ |
|---|---|---|---|---|
| 0.001 | -0.443166 | 0.01119 | -0.402010 | 0.02078 |
| 0.002 | -0.443196 | 0.02064 | -0.402094 | 0.03465 |
| 0.005 | -0.443333 | 0.04116 | -0.402423 | 0.06243 |
| 0.010 | -0.443655 | 0.06374 | -0.403121 | 0.09883 |
| 0.020 | -0.444537 | 0.10982 | -0.404844 | 0.15698 |
| 0.050 | -0.448374 | 0.21427 | -0.411407 | 0.28984 |
| 0.100 | -0.457246 | 0.35707 | -0.424741 | 0.46237 |
| 1/3 | -0.517954 | 0.87661 | -0.501575 | 1.05382 |

Figure 2 shows a log-log plot of $\varepsilon(\delta, J_2) - \varepsilon(0, J_2)$ vs. $\delta$ for $J_2 = 0$ (HAF) and $J_{2c}$. The scaling exponent of $\beta = 4/3$ holds at $J_{2c}$ up to at least $\delta = 0.10$. The HAF curve yields $\beta = 1.450$, distinctly less than $2\alpha = 1.50$. The power law fit of the energy gain is more accurate than the scaling expression with logarithmic corrections. Least square fits to Eq. (2) for $\ln \Delta$ vs $\ln \delta$ and $\ln |\varepsilon(\delta) - \varepsilon(0)|$ vs $\ln \delta$ up to $\delta = 0.10$ lead to the exponents $\alpha$, $\beta$ and coefficients A, B in Table 2. Increasing $J_2$ decreases the coefficient of the marginal operator that is responsible for logarithmic corrections. Indeed, $\alpha$ and $\beta$ decrease smoothly to the scaling values at $J_{2c}$, while A and B increase slightly. We also observe that the error bars decrease as $J_2$ increases from 0 to $J_{2c}$. The $J_{2c}$ results agree with theory within our improved numerical accuracy up to at least $\delta = 0.10$, which is well into the regime of spin-Peierls systems.[22-24] As previously noted,[2,4,20] the ST gap at $J_2 = 0$ is more accurately represented by Eq. (2) than by scaling theory with logarithmic corrections.

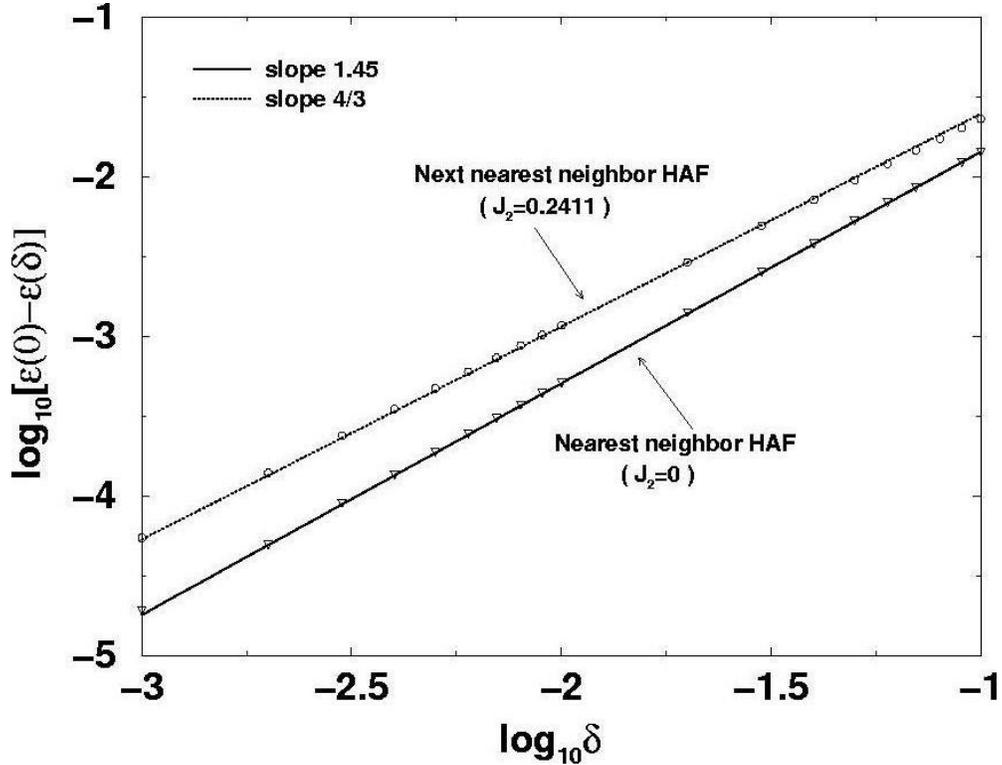

**Fig. 2** Log-log plot of energy per site versus dimerization at $J_2 = J_{2c} = 0.2411$ and $J_2 = 0$. The slope at $J_{2c}$ is 4/3 as predicted by scaling theory.

Our DMRG results show that the power laws in Eq. (2) accurately fit the ST gap and energy stabilization on dimerization of $H(\delta, J_2)$ in the interval $0.001 \leq \delta \leq 0.10$ for any $J_2 \leq J_{2c}$. The HAF and scaling analysis in Ref. 2 extends to far smaller $\delta$ and to very low temperature. Such mathematical properties of spin chains are largely beyond experimental comparisons. In terms of physical realizations, spin-Peierls systems[22-24] with strong coupling to the lattice and $\delta \sim 0.10$ at low T are required in order to minimize competition from second neighbor or interchain exchange or from anisotropic or antisymmetric corrections to J.

**Table 2** Fit to Eq. (2) of the ST gap $\Delta(\delta,J_2)$ and energy $\varepsilon(\delta,J_2)$ for dimerization $0.001 \leq \delta \leq 0.10$ and frustration $J_2 \leq J_{2c}$. The standard deviation ($\sigma$) for the parameters is also shown.

|  | $\Delta(\delta,J_2) = A\delta^\alpha$ | | | | $\varepsilon(\delta,J_2) - \varepsilon(0,J_2) = -B\delta^\beta$ | | | |
| --- | --- | --- | --- | --- | --- | --- | --- | --- |
| $J_2$ | $\alpha$ | $\sigma_\alpha$ | A | $\sigma_A$ | $\beta$ | $\sigma_\beta$ | B | $\sigma_B$ |
| 0.000 | 0.7475 | 0.0075 | 2.0375 | 0.0672 | 1.4417 | 0.0018 | 0.3891 | 0.0031 |
| 0.050 | 0.7365 | 0.0066 | 2.0720 | 0.0606 | 1.4388 | 0.0026 | 0.4331 | 0.0051 |
| 0.100 | 0.7246 | 0.0052 | 2.1090 | 0.0492 | 1.4137 | 0.0023 | 0.4497 | 0.0047 |
| 0.150 | 0.7101 | 0.0041 | 2.1431 | 0.0394 | 1.3769 | 0.0014 | 0.4527 | 0.0029 |
| 0.200 | 0.6903 | 0.0027 | 2.1585 | 0.0261 | 1.3469 | 0.0028 | 0.4735 | 0.0059 |
| 0.2411 | 0.6715 | 0.0018 | 2.1709 | 0.0172 | 1.3078 | 0.0027 | 0.4736 | 0.0056 |

The energy stabilization of the HAF on dimerization drives the spin-Peierls transition. In usual approximations of linear spin-phonon coupling, $\gamma = (dJ/du)_0$ for displacements $u(-1)^n$, and a harmonic lattice with force constant K, the equilibrium dimerization $\delta(0)$ at T = 0 is given by[2]

$$\frac{KJ}{\gamma^2} = -\frac{1}{\delta(0)}\left(\frac{\partial \varepsilon}{\partial \delta}\right)_{\delta(0)} = B\beta\delta(0)^{\beta-2} \qquad (3)$$

Strong coupling $\gamma$ and/or small K result in large dimerization. The divergence as $\delta^{-0.55}$ is rather weak in spin-Peierls systems with $\delta(0) > 0.05$. A large ST gap localizes the GS and leads to rapid size convergence for the function $\delta^{-1}\varepsilon'(\delta)$. Direct solution[24] of $H(\delta,0)$ for 22 spins in Eq. (1) agrees within a few percent for $\delta = 0.05$ with the DMRG analysis and rapidly becomes more quantitative with increasing $\delta$. The large ST gap in the dimerized singlet ground state localizes the wave function and speeds up size convergence.

Our results for $H(\delta,J_2)$ for the special case of $J_{2c}$ and for experimentally realized range of dimerization, agrees quantitatively with scaling theory both for the ST gap and the stabilization energy (with exponents of 2/3 and 4/3, respectively) up to at least $\delta \sim 0.2$. Indeed, direct least squares fits of the gap and energy per site vs. dimerization (instead of linear fits to the log-log plot) give exponents of 0.667 and 1.333 with standard deviations of 0.0014 for the gap and 0.00033 for the energy per site. Logarithmic corrections are expected for $J_2 < J_{2c}$. Power laws with exponents different from the universal values of 2/3 and 4/3, as found numerically in Table 2 for the ST gap and stabilization energy, are then more accurate than the scaling theory expressions with logarithmic corrections. Thus the logarithmic corrections complicate the scaling theory expressions for the ST gap and stabilization energy of the linear HAF, thereby restricting universal power law behavior to very small values of $\delta$.

**Acknowledgements**: ZGS thanks Jawaharlal Nehru Centre for Advanced Scientific Research for a visiting professorship. SR thanks Indo-French Centre for Promotion of Advanced Research for a research grant under Project 3108-3. DS thanks the Department of Science and Technology, India for financial support under the project SP/S2/M-11/2000.